\documentclass[showpacs,twocolumn,aps,prl,superscriptaddress]{revtex4}
\usepackage{amsmath,amssymb}
\usepackage{graphicx}
\usepackage{dcolumn}
\usepackage{color}
\usepackage{epsfig}
\usepackage{epstopdf}
\usepackage{subfigure}
\usepackage{amsfonts}
\usepackage{xcolor}
\usepackage{float}

\usepackage{bm}
\newcommand{\tr}[1]{\,\mathrm{tr}\left\lbrace  #1 \right\rbrace}
\newcommand{\ket}[1]{\left\vert#1\right\rangle}
\newcommand{\bra}[1]{\left\langle#1\right\vert}

\newcommand\lavg{\left\langle}
\newcommand\ravg{\right\rangle}
\newcommand\be{\begin{equation}}
\newcommand\ee{\end{equation}}

\let\veps=\varepsilon

\begin{document}

\title{Irreversible work and inner friction in quantum thermodynamic processes}

\author{F.~Plastina}
\affiliation{Dip.  Fisica, Universit\`a della Calabria, 87036
Arcavacata di Rende (CS), Italy} \affiliation{INFN - Gruppo
collegato di Cosenza, Cosenza Italy}
\author{A. Alecce}
\affiliation{Dipartimento di Fisica e Astronomia "G. Galilei",
Universit\`{a} degli Studi di Padova, via Marzolo 8, 35131 Padova
(Italy)}
\author{T. J. G. Apollaro}
\affiliation{Dip.  Fisica, Universit\`a della Calabria, 87036
Arcavacata di Rende (CS), Italy} \affiliation{INFN - Gruppo
collegato di Cosenza, Cosenza Italy}
\author{G. Falcone}
\affiliation{Dip.  Fisica, Universit\`a della Calabria, 87036
Arcavacata di Rende (CS), Italy} \affiliation{INFN - Gruppo
collegato di Cosenza, Cosenza Italy}
\author{G. Francica}
\affiliation{Dip.  Fisica, Universit\`a della Calabria, 87036
Arcavacata di Rende (CS), Italy} \affiliation{INFN - Gruppo
collegato di Cosenza, Cosenza Italy}
\author{F.~Galve}
\affiliation{ IFISC (UIB-CSIC), Instituto de F\'isica
Interdisciplinar y Sistemas Complejos, UIB Campus, E-07122 Palma
de Mallorca, Spain}
\author{N. Lo Gullo}
\affiliation{Dipartimento di Fisica e Astronomia "G. Galilei",
Universit\`{a} degli Studi di Padova, via Marzolo 8, 35131 Padova
(Italy)} \affiliation{CNISM, Sezione di Padova, Italy}
\author{R. Zambrini}
\affiliation{ IFISC (UIB-CSIC), Instituto de F\'isica
Interdisciplinar y Sistemas Complejos, UIB Campus, E-07122 Palma
de Mallorca, Spain}

\begin{abstract}
We discuss the thermodynamics of closed quantum systems driven out
of equilibrium by a change in a control parameter and undergoing a
unitary process. We compare the work actually done on the system
with the one that would be performed along ideal adiabatic and
isothermal transformations. The comparison with the latter leads
to the introduction of irreversible work, while that with the
former leads to the introduction of inner friction. We show that
these two quantities can be treated on equal footing, as both can
be linked with the heat exchanged in thermalization processes and
both can be expressed as relative entropies. Furthermore, we show
that a specific fluctuation relation for the entropy production
associated with the inner friction exists, which allows the inner
friction to be written in terms of its cumulants.

\end{abstract}
\maketitle

With the increasing ability to manufacture and control microscopic
systems, we are approaching the limit where quantum fluctuations,
as well as thermal ones, become important when trying to put
nanomachines and quantum engines to useful purposes
\cite{seifert,seeeg}. To discuss engines performances, e.g. for
heat-to-work conversion, one typically starts by considering
reversible transformations that drive the system from an
equilibrium configuration to another one. However, if the system
is pushed faster than the thermalization time, such
transformations are irreversible, and can lead outside the
manifold of equilibrium states \cite{espo,revfluctrel,rev2}.
Nonetheless, these processes are of interest as the reversible
protocols, despite enjoying very good efficiencies, give rise to
very small output powers \cite{alle}. The irreversibility of a
process is hence related both to better performances and to lack
of control, leading to entropy production \cite{entropro}.

To analyze irreversibility and entropy production in the quantum
realm, we consider a system initially kept in equilibrium and
subject to a finite time adiabatic transformation. While its
initial state is prepared by keeping it in contact  with a thermal
bath, the system is then thermally isolated and subject to a
parametric change of its Hamiltonian from an initial
$H_i=H[\lambda_i]$ to a final $H_f=H[\lambda_f]$ in a finite time
$\tau$. The process is defined by the time variation of the work
parameter $\lambda(t)$, changing from $\lambda(t=0) = \lambda_i$
to $\lambda(\tau) = \lambda_f$.
\begin{figure}
\centering{\includegraphics[width=0.9\columnwidth]{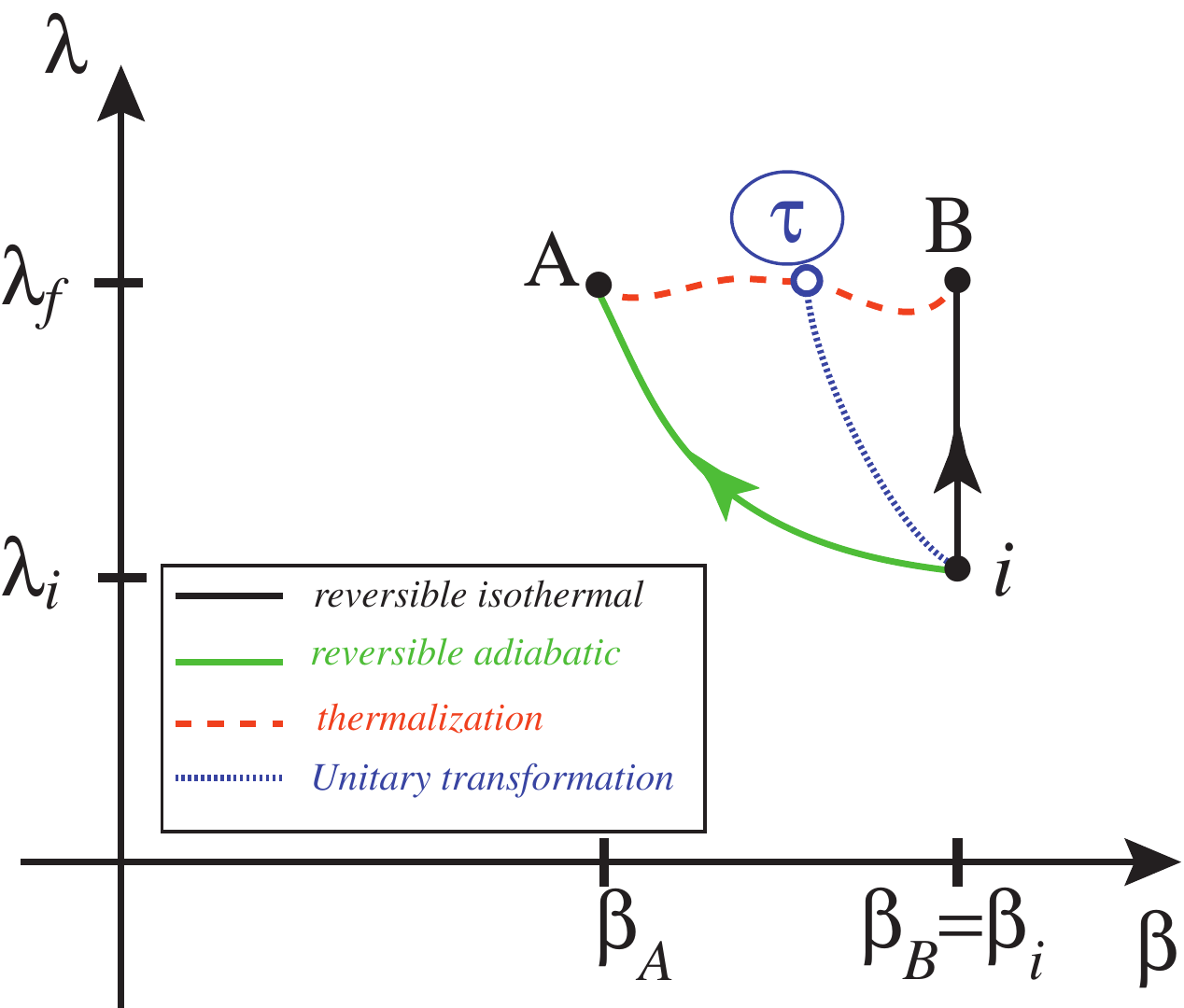}}
\caption{(Color online) Sketch of the transformations considered.
The full black circles represent equilibrium states:
$\rho_i=Z^{-1}_i e^{-\beta_i H_i}$ $\rho_A=Z^{-1}_A e^{-\beta_A
H_f}$, and $\rho_B=Z^{-1}_B e^{-\beta_B H_f}$. The empty blue one,
instead, is the state at time $\tau$, $\rho_{\tau} = U(\tau,0)
\rho_i U^{\dag}(\tau,0)$. } \label{trasforma}
\end{figure}

The work $w$ performed on the system during such a process is a
stochastic variable with an associated probability density $p(w)$
\cite{jarz,mukamel,revfluctrel}, which can be reconstructed
experimentally \cite{pek,recon} through the characteristic
function \cite{talkner}. The fluctuations of work are constrained
by the Jarzynski relation \cite{jarz} \be \lavg e^{-\beta_i w}
\ravg = e^{-\beta_i \Delta F} \, , \label{jeq}\ee where
$\beta_i\equiv\beta_B$ is the initial inverse temperature, while
$\Delta F= F[\lambda_f, \beta_B] - F[\lambda_i, \beta_i]$ is the
free energy difference between two equilibrium configurations (at
the same temperature) corresponding to the initial and final
Hamiltonian. This fluctuation relation encodes the full nonlinear
response of a system to a time-dependent change of its
Hamiltonian. Through the use of the Jensen's inequality, this
relation implies that $\lavg w \ravg \geq \Delta F$. This, in
turn, leads to the introduction of the so called average
irreversible work \cite{crooks}, $\lavg w_{irr} \ravg = \lavg w
\ravg - \Delta F \geq 0$ . Loosely speaking, $\lavg w_{irr} \ravg$
gives a measure of the irreversibility introduced in performing
the unitary transformation $U(\tau,0)$ generated by the
Hamiltonian $H[\lambda(t)]$ between $t=0$ and $t=\tau$.

The situation is sketched in Fig. (\ref{trasforma}), where the
point $i$ corresponds to the initial state $\rho_i = e^{-\beta_i
H_i}/Z[\lambda_i, \beta_i]$, while the point $\tau$ corresponds to
the state $\rho_{\tau} = U(\tau,0) \rho_i U^{\dag}(\tau,0)$.
Notice that this latter point does not lay on the manifold of
equilibrium states (it could do so only in the limit $\tau
\rightarrow \infty$, in which the transformation would become
quasi-static).

It has been shown in Ref. \cite{defflutz} that $\lavg w_{irr}
\ravg$ is given by the distance between the actual final state
$\rho_{\tau}$ and the (hypothetical) equilibrium state $\rho_B =
e^{-\beta_B H_f}/Z(B)$, evaluated through the quantum relative
entropy (or quantum Kullback-Leibler divergence) \be \lavg w_{irr}
\ravg = \frac{1}{\beta_B} \, D(\rho_{\tau} || \rho_B) \, .
\label{istlutz} \ee The irreversible work has been used to learn
about the amount of irreversibility of a given process in a
variety of cases, ranging from simple harmonic systems
\cite{carlisle}, to spin chains \cite{dorner,campap}, and
ultra-cold gases \cite{sindona}.

By using the definition of free energy, $F = {\cal U} - TS$, where
${\cal U} = \tr{\rho H}$ is the internal energy, while $S$ denotes
the thermodynamic entropy (here evaluated as $S= -\tr{\rho \, \ln
\rho}$ for equilibrium states, at points $i$, $A$ and $B$), one
can show that $\lavg w_{irr} \ravg$ is related to the (average)
heat required to let the system thermalize, starting from the
state $\rho_{\tau}$, by keeping it in contact with a heat bath at
temperature $T_B = \beta_B^{-1}$ (see also \cite{campap} ): \be
\lavg w_{irr} \ravg = T_B (S_{B} - S_i) - \lavg Q^{th}_{\tau
\rightarrow B} \ravg \, , \label{qth} \ee where $\lavg
Q^{th}_{\tau \rightarrow B} \ravg = \tr{(\rho_B -\rho_{\tau})
H_f}$ is the energy taken by the system in the thermalization
process leading it from the state $\rho_{\tau}$ to $\rho_B$
\cite{footnote1}. Thus, the irreversible work is both related to
{\it i}) energy flow (heat) and entropy change when returning to
the isothermal branch $i\rightarrow B$ we had left because of the
unitary driving $U(\tau,0)$, and {\it ii}) to the distance between
the equilibrium state $\rho_B$ and the actual one, $\rho_\tau$.

{\it Inner friction.-}  On the other hand, in the context of
finite time thermodynamics, one is often led to compare the $i
\rightarrow \tau$ process with the reversible adiabatic
transformation $i \rightarrow A$ (rather than with the isothermal
$i \rightarrow B$). Indeed, adiabatic transformations enter the
Carnot and the Otto cycles and have been, therefore, largely
studied and discussed  so far \cite{adiabs}. In particular, when
analyzing finite time adiabatic transformations, performed on
thermally isolated quantum systems, it is quite natural to
introduce the concept of {\it inner friction}, defined as the
difference between the actual work performed on the system and the
ideal one, done along an ideal reversible adiabatic transformation
\cite{kosloff00,kosloffMANY,wang12,wang}. This difference comes in
when the system is unable to adiabatically follow the control
protocol, typically because of some (inner or intrinsic) degrees
of freedom do not commute with the control Hamiltonian.

As is the case for the irreversible work, we will show that inner
friction too is related to a distance between the states attained
through the hypothetical reversible and the actual unitary
transformations, respectively. Furthermore, we will show that
inner friction (and, in particular, the entropy production
associated to it) can be described through a stochastic variable
fulfilling a thermodynamic fluctuation relation.

Explicitly, in a reversible and quasi-static adiabatic
transformation, the energy levels of the system experience a
change as $H[\lambda_i]$ is slowly modified into $H[\lambda_f]$,
but the occupation probabilities of these levels stay the same (we
assume that no level crossing occurs), so that, for every
eigenstate $\ket{\veps_m^{(f)}}$ of $H_f$, the (constant)
population is still given by the initial value $P_m^{(i)}=\exp
\{-\beta_i \veps_m^{(i)} \} / Z[\lambda_i, \beta_i]$. If the
energy eigenvalue is changed from $\veps_m^{(i)}$ to
$\veps_m^{(f)}$ as $\lambda(t)$ goes from $\lambda_i$ to
$\lambda_f$, this implies that also temperature has changed. Its
final value is such that $P_m^{(i)}= P_m^{(A)}$; that is $\exp
\{-\beta_i \veps_m^{(i)} \} / Z[\lambda_i, \beta_i] = \exp
\{-\beta_A \veps_m^{(f)} \} / Z[\lambda_f, \beta_A]$. The
requirements for a reversible adiabatic process are indeed very
tight as this relation has to hold for any adiabatically evolved
eigenstate; that is, for every $m$. This, in turn, implies that
all energy gaps of the system have to change by the ratio
$\beta_i/\beta_A$ \cite{nori}.

Under such conditions, the work performed on the $i\rightarrow A$
transformation is given by \be\lavg w_{i \rightarrow A} \ravg =
{\cal U}_{A} - {\cal U}_{i} \equiv \sum_m P_m^{(i)} (\veps_m^{(f)}
- \veps_m^{(i)}) \, \label{servedu} \ee This is, once again,
different from the average work performed during the actual
(unitary and finite-time) process $i \rightarrow \tau$. The
difference between the two, \be \lavg w_{fric} \ravg = \lavg w
\ravg - \lavg w_{i \rightarrow A} \ravg \label{infr}\ee has been
called inner friction as it is due to unwanted transitions that
one would typically associate with heat. Indeed, by its definition
and as discussed in details below, the inner friction is precisely
the `excess heat' the system has taken and that it would release
to the environment if thermalizing at inverse temperature
$\beta_A$.

We now show that, similarly to Eq. (\ref{istlutz}) for the
irreversible work, $\lavg w_{fric} \ravg$ is given by the distance
between the actual final state $\rho_{\tau}$ and the
(hypothetical) equilibrium state $\rho_A = e^{-\beta_A H_f}/Z(A)$,
evaluated through the quantum relative entropy \be \lavg w_{fric}
\ravg = \frac{1}{\beta_A} \, D(\rho_{\tau} || \rho_A)
\label{nuovadist}\ee Indeed, by eq. (\ref{servedu}), one gets
\begin{eqnarray}
\lavg w_{fric} \ravg & = & \lavg w\ravg - \lavg w_{i \rightarrow
A} \ravg  =  \tr{\rho_{\tau} H_f} -
{\cal U}_A = \nonumber \\
&=& \sum_m \veps_m^{(f)} \left [ \bra{\veps_m^{(f)} }\rho_{\tau}
\ket{\veps_m^{(f)}} - P_m^{(A)} \right ] \, , \label{wfr}
\end{eqnarray} while
\begin{eqnarray}
D(\rho_{\tau} || \rho_A) &=& \tr{\rho_{\tau} \ln \rho_{\tau}} -
\tr{\rho_{\tau} \ln \rho_{A}} = \nonumber \\
&=& \sum_m P_m^{(i)} \, \ln  P_m^{(i)} -
\bra{\veps_m^{(f)} }\rho_{\tau} \ket{\veps_m^{(f)}} \ln P_m^{(A)} = \nonumber \\
&=& \sum_m \ln P_m^{(A)} \left [ P_m^{(i)} -
\bra{\veps_m^{(f)} }\rho_{\tau} \ket{\veps_m^{(f)}} \right ] = \nonumber \\
&=& \sum_m \beta_A \veps_m^{(f)} \left [ \bra{\veps_m^{(f)}
}\rho_{\tau} \ket{\veps_m^{(f)}} - P_m^{(A)} \right ] \, ,\label{distance}
\end{eqnarray}
where we used $P_m^{(i)}=P_m^{(A)}$. These two relations, taken
together, demonstrate Eq. (\ref{nuovadist}).

Being given by a relative entropy, $\lavg w_{fric} \ravg$ is thus
always greater than zero by the Klein's inequality, \cite{donor}.

Furthermore, similarly to what has been done in Ref.
\cite{defflutz}, one can find a better (geometric) lower bound
expressed in terms of the finite Bures length: \be \beta_A \lavg
w_{fric} \ravg \geq \frac{8}{\pi^2} \, {\cal L}^2(\rho_{\tau},
\rho_A) \, , \label{noncapisco} \ee where, for any two density
operators, ${\cal L}$ is given in terms of the fidelity ${\cal F}$
between those states, ${\cal L}(\rho_1,\rho_2) =
\arccos\{\sqrt{F(\rho_1,\rho_2)} \, \}$, with $${\cal
F}(\rho_1,\rho_2) = \left [ \tr { \sqrt{\sqrt{\rho_1} \, \rho_2 \,
\sqrt{\rho_1}} } \right ]^2 \, .$$ The inner friction is hence
bounded from below by the geometric distance between the actual
density operator $\rho_{\tau}$ at the end of the process and the
corresponding equilibrium operator $\rho_A$, as measured by the
Bures angle ${\cal L}$.

This gives a precise meaning to the idea
that, when performing an adiabatic transformation in a finite
time, the amount of work that `gets lost' is larger when the
system is brought far and far away from equilibrium.

Going back to the thermalization process $\tau \rightarrow A$, we
have that the average heat taken by the system to thermalize at
$T_A=\beta_A^{-1}$ is given by \be \beta_A \lavg Q^{th}_{\tau
\rightarrow A} \ravg  = - \beta_A \lavg w_{fric} \ravg = -
D(\rho_{\tau} || \rho_A) \, ,\ee which easily compares with the
analogous expression  for $\lavg Q^{th}_{\tau \rightarrow A}
\ravg$ reported in Eq. (\ref{qth}), as $S_i = S_A$ for an
adiabatic process.

The heat exchange in a thermalization process is a quantity of
fundamental interest as, through the Landauer principle, it is
linked to information processing, storing and erasing protocols,
as well as the information-to-energy conversion, \cite{toy}.
Indeed, attention has been given to this subject extensively in
the literature \cite{john} as any attempt at exploring the
fundamental energetic limits of information processing would need
to measure such an heat.

Comparing the definitions of the two average heat exchanges, one
obtains $ \lavg Q^{th}_{\tau \rightarrow A} \ravg - \lavg
Q^{th}_{\tau \rightarrow B} \ravg = {\cal U}_A - {\cal U}_B$,
which gives an explicit relation between irreversible work and
inner friction: \be \lavg w_{irr} \ravg - \lavg w_{fric} \ravg =
({\cal U}_A - {\cal U}_B) - T_i(S_A - S_B) \, , \label{difirfr}\ee
or, stated differently, $ \lavg w_{irr} \ravg + F_B + T_B S_i =
\lavg w_{fric} \ravg + F_A + T_A S_i \, ,$ \cite{footnote2}.

{\it Fluctuation relation.-} For a reversible and infinitely slow
$i \rightarrow \tau$ process, the actual final state $\rho_{\tau}$
would coincide with the equilibrium state $\rho_A$, with no net
entropy change, as $S_i = S_A$. This latter equality implies that
$\beta_A {\cal U}_A - \beta_i {\cal U}_i = \beta_A F_A - \beta_i
F_i$.

In the actual, finite time process, instead, the entropy
production is non-zero, as un-wanted transitions between adiabatic
energy eigenstates may occur, as signalled by $\lavg
w_{fric}\ravg$. We can fully characterize the entropy production
due to these {\it diabatic} transitions by defining an auxiliary
entropic variable $s$, obtained (as by now usual) by a
two-measurement protocol in which energy is measured at the
beginning and at the end of the process. Given the two outcomes
(say $\veps^{(i)}_n$ and $\veps_m^{(f)}$, respectively), we can
build the stochastic variable $$s = \beta_A \veps_m^{(f)} -
\beta_i \veps_n^{(i)} \, ,$$ which is distributed according to the
probability density \be p(s) = \sum_{n,m}P_n^{(i)} \,
P_{n\rightarrow m}^{(\tau)} \, \delta (s - \beta_A \veps_m^{(f)} +
\beta_i \veps_n^{(i)}) \, ,\ee with $P_n^{(i)} = Z_i^{-1}
e^{-\beta_i \veps_n^{(i)}}$ and $P_{n\rightarrow m}^{(\tau)} =
\left |\bra{\veps_m^{(f)}} U(\tau,0) \ket{\veps_n^{(i)}} \right
|^2$.

The average value of $s$ gives $\lavg s \ravg = \beta_A
\tr{\rho_{\tau} H_f }- \beta_i {\cal U}_i$, which, for a
reversible quasi-static transformation, with $\rho_{\tau} \equiv
\rho_A$, would give the sought combination of internal energies:
$\beta_A {\cal U}_A - \beta_i {\cal U}_i$. Furthermore, a
fluctuation relation similar to  Eq. (\ref{jeq}), can be obtained:
\begin{eqnarray}
\lavg e^{-s} \ravg &=& \sum_{n,m} P_n^{(i)} \, P_{n\rightarrow
m}^{(\tau)} \, e^{- (\beta_A \veps_m^{(f)} - \beta_i
\veps_n^{(i)})}
= \nonumber \\
&=& \frac{Z_A}{Z_i} \equiv e^{-(\beta_A F_A - \beta_i F_i)}
\end{eqnarray}
This is a special case of a more general relation derived by
Tasaki \cite{tasaki}, which is of particular relevance here due to
its relation with $\lavg w_{fric} \ravg$. Indeed, by use of
Jensen's inequality, this implies that \be \lavg s \ravg \geq
\beta_A F_A - \beta_i F_i \ee which shows that the average entropy
production in the actual process is always larger than zero
$$\lavg \Sigma \ravg := \lavg s \ravg - (\beta_A F_A - \beta_i
F_i) \geq 0 \, .$$ This latter quantity is easily shown to be
related to the inner friction and to the corresponding relative
entropy \be \lavg \Sigma \ravg \equiv \beta_A \lavg w_{fric} \ravg
\equiv D(\rho_{\tau}|| \rho_A) \, .\ee

In analogy to what has been done by Jarzynski in Ref. \cite{jarz},
where the cumulants of the distribution of the work done in the
process $\lambda_i \rightarrow \lambda_f$ have been related to the
free energy difference $F_B - F_i$, we can show that the cumulants
$C_n $ of the distribution of the variable $s$ are related to the
combination of free energies $\beta_A F_A - \beta_i F_i$ as
\cite{footnote3} \be - ( \beta_A F_A - \beta_i F_i) = \sum_{n=1}
\frac{(-1)^n}{n!} C_n \, . \label{cumu}\ee Finally, this implies
that the inner friction can be expressed as a combination of the
cumulants of order larger than $2$ \be \lavg \Sigma \ravg =
\beta_A \lavg w_{fric} \ravg = \frac{C_2}{2} - \frac{C_3}{6} +
\ldots \, ,\ee where $C_2 = \lavg s^2 \ravg - \lavg s \ravg^2$ is
the variance, $C_3 = \lavg s^3\ravg -3 \lavg s^2 \ravg  \lavg s
\ravg + 2 \lavg s \ravg^3$ is the skewness and so on.

{\it Discussion.-} We have shown that it is meaningful to consider
the closeness to an ideal adiabatic transformation of an actual
unitary evolution of a generic quantum system, driven out of
equilibrium by changing in time a work parameter $\lambda$
entering its Hamiltonian. The comparison of the work done on the
system in the two cases naturally leads to the concept of inner
friction, which is related to the heat the system would release to
a thermal bath, if thermalizing at the final temperature
$\beta_A^{-1}$. Inner friction also comes out naturally when
considering the statistics of the entropy irreversibly produced
during the actual process, which satisfies a fluctuation relation
analogous to the Jarzynski equality. Indeed, the average excess
entropy, produced due to the irreversible nature of the actual
process, precisely coincides with the inner friction (divided by
the final temperature), which, therefore, can be expressed as a
cumulant series.

Inner friction has been previously considered in the literature
through simple microscopic models of thermal engines, where the
working substance is composed of interacting spin-dimers
\cite{kosloffMANY} or of an harmonic oscillator \cite{kosloffOSC}.
In the former, the friction comes from interaction; in the latter
it comes from the intrinsic non-commutativity at different times
of the oscillator Hamiltonian whose frequency is being varied.
Further irreversible sources of noise/friction can be added
artificially to the oscillator, such as frequency- or phase-noise
\cite{kosloffOSC2}.

Strategies against inner friction have been considered mainly
under the generic names `shortcuts to
adiabaticity'\cite{shortcuts} (where control sequences
$\lambda(t)$ are designed such that the irreversibility at the end
of the adiabatic branch is minimized) and `quantum lubrication'
\cite{kosloffLUBR} (where the coherences of $\rho_\tau$ in the
$|\epsilon^{(f)}\rangle$ basis are minimized through an additional
dephasing noise, thus minimizing $D(\rho_\tau||\rho_A)$ see
eq.(\ref{distance})).

It must be stressed though, that to the best of our knowledge, the
irreversibility caused by inner friction had never been associated
to a distance from an equilibrium state, nor a to any fluctuation
theorem.

As a final remark, we would like to emphasize once again the
assumption on which our treatment relies, namely the definition of
$\beta_A$. The quantum adiabatic theorem \cite{messiah} guarantees
that, in the absence of level crossings, a very slow
transformation would lead from the initial state $\rho_i =
Z_i^{-1} \sum_n e^{-\beta_i \veps_n^{(i)}} \, \ket{\veps_n^{(i)}}
\bra{\veps_n^{(i)}}$ to a final state with the same population and
new eigenstates, $\rho_A =Z_i^{-1} \sum_n e^{-\beta_i
\veps_n^{(i)}} \, \ket{\veps_n^{(f)}} \bra{\veps_n^{(f)}}$. We
assumed this state to be a thermal equilibrium one at inverse
temperature $\beta_A$. As mentioned above, this is a tight
requirement that cannot always be fulfilled. There are, however,
relevant cases in which there is no such a problem: that of an
harmonic oscillator (or of an harmonically trapped gas) and that
of a two level system (or, more generally, a collection of
non-interacting spins) whose frequency is parametrically changed
during the process. For these systems, $\beta_A$ can always be
defined, as well as for any other quantum system undergoing a
transformation for which all of the initial energy gaps scale by
the same factor. In all of these cases, our analysis is meaningful
and the comparison of an actual unitary evolution with a
reversible and quasi-static adiabatic transformation is well
defined.

\vskip 12pt {\bf Acknowledgement} All the authors acknowledge
support from COST MP1209 Action. FP, GF and NLG acknowledge
insightful discussions with Michele Campisi, John Goold and Mauro
Paternostro. T.J.G.A. is supported by the European Commission, the
European Social Fund, and the Region Calabria through the program
POR Calabria FSE 2007-2013-Asse IV Capitale Umano-Obiettivo
Operativo M2.  F.G. and R.Z. acknowledge MINECO, CSIC, the EU
commission, and FEDER funding under Grants No. FIS2007-60327
(FISICOS), No. FIS2011-23526 (TIQS), postdoctoral JAE program
(ESF).


\begin{thebibliography}{}
\bibitem{seifert}
U. Seifert, Rep. Prog. Phys. {\bf 75}, 126001 (2012).

\bibitem{seeeg}
See, e.g., the experimental proposal in C. Bergenfeldt, P.
Samuelsson, B. Sothmann, C. Flindt, and M. B\"{u}ttiker, Phys.
Rev. Lett. {\bf 112}, 076803 (2014).

\bibitem{espo}
M. Esposito, U. Harbola, and S. Mukamel, Rev. Mod. Phys. {\bf 81},
1665 (2009).

\bibitem{revfluctrel}
C. Jarzynski, Annu. Rev. Condens. Matter Phys. {\bf 2}, 329
(2011); M. Campisi, P. H\"anggi, and P. Talkner, Rev. Mod. Phys.
{\bf 83}, 771 (2011).

\bibitem{rev2}
A. Polkovnikov, K. Sengupta, A. Silva, and M. Vengalattore Rev.
Mod. Phys. {\bf 83}, 863 (2011).

\bibitem{alle}
A. E. Allahverdyan, R. S. Johal, and G. Mahler, Phys. Rev. E {\bf
77}, 041118 (2008); M. Esposito, R. Kawai, K. Lindenberg, and C.
Van den Broeck, Phys. Rev. Lett. {\bf 105}, 150603 (2010); G.
Benenti, K. Saito, and G. Casati, Phys. Rev. Lett. {\bf 106},
230602 (2011); U. Seifert, Phys. Rev. Lett. {\bf 106}, 020601
(2011); K. Brandner, K. Saito, and U. Seifert, Phys. Rev. Lett.
{\bf 110}, 070603 (2013); A. E. Allahverdyan, K. V. Hovhannisyan,
A. V. Melkikh, and S. G. Gevorkian, Phys. Rev. Lett. {\bf 111},
050601 (2013).

\bibitem{entropro}
M. Esposito and C. Van den Broeck, Phys. Rev. Lett. {\bf 104},
090601 (2010); S. Ciliberto, A. Imparato, A. Naert, and M. Tanase,
Phys. Rev. Lett. {\bf 110}, 180601 (2013); J. M. Horowitz and J.
M. R. Parrondo, New J. Phys. {\bf 15}, 085028 (2013).

\bibitem{jarz}
C. Jarzynski, Phys. Rev. Lett. {\bf 78}, 2690 (1997).

\bibitem{mukamel}
J. Kurchan, arXiv:cond-mat/0007360v2 (2000); S. Mukamel, Phys.
Rev. Lett. {\bf 90}, 170604 (2003).

\bibitem{pek}
J. P. Pekola, P. Solinas, A. Shnirman, and D. V. Averin,
arXiv:1212.5808 (2012).

\bibitem{recon}
R. Dorner, S. R. Clark, L. Heaney, R. Fazio, J. Goold, V. Vedral,
Phys. Rev. Lett. {\bf 110}, 230601 (2013); L. Mazzola, G. De
Chiara, and M. Paternostro, Phys. Rev. Lett. {\bf 110}, 230602
(2013); T. Batalh\~{a}o {\it et al}., arXiv:1308.3241 (2013); M.
Campisi, R. Blattmann, S. Kohler, D. Zueco, and P. H\"{a}nggi, New
J. Phys. {\bf 15}, 105028 (2013).

\bibitem{talkner}
P. Talkner, E. Lutz and P. H\"{a}nggi, Phys. Rev. E {\bf 75},
050102(R) (2007); P. Talkner and P. H\"{a}nggi, J. Phys. A: Math.
Theor. {\bf 40} F569 (2007).

\bibitem{crooks}
G. E. Crooks, Phys. Rev. E {\bf 60}, 2721 (1999).

\bibitem{defflutz}
S. Deffner and E. Lutz, Phys. Rev. Lett. {\bf 105}, 170402 (2010).

\bibitem{carlisle}
F. Galve and E. Lutz, Phys. Rev. A {\bf 79}, 032327 (2009); F.
Galve and E. Lutz, Phys. Rev. A {\bf 79}, 055804 (2009); Carlisle
et al., arXiv:1403.0629 (2014).

\bibitem{dorner}
A. Silva, Phys. Rev. Lett. {\bf 101}, 120603 (2008); F.N.C. Paraan
and A. Silva, Phys. Rev. E {\bf 80}, 061130 (2009); R. Dorner, J.
Goold, C. Cormick, M. Paternostro, V. Vedral, Phys. Rev. Lett.
{\bf 109},160601 (2012).

\bibitem{campap}
T. J. G. Apollaro, G. Francica, M. Paternostro, M. Campisi,
arXiv:1406.0648 (2014).

\bibitem{sindona}
J. Yi, Y. W. Kim, P. Talkner, Phys. Rev. E {\bf 85}, 051107
(2012); A. Sindona, J. Goold, N. Lo Gullo, F. Plastina, New J.
Phys. {\bf 16}, 045013(2014).

\bibitem{footnote1} This relation comes directly from the
definition of work, $\lavg w \ravg = \tr{\rho_{\tau} H_f} - {\cal
U}_i$, which implies $$\lavg w \ravg + \lavg Q^{th}_{\tau
\rightarrow B} \ravg = {\cal U}_{B} - {\cal U}_{i}\equiv F_{B} -
F_i + T_B (S_{B} - S_{i})\, .$$ This is equivalent to Eq.
(\ref{qth}), which, together with Eq. (\ref{istlutz}) gives $
\beta_B \lavg Q_{\tau\rightarrow B}^{th} \ravg = S_B - S_i -
D(\rho_{\tau} || \rho_B)$.

\bibitem{adiabs}
M. J. Henrich , G. Mahler and M. Michel, Phys. Rev. E {\bf 75},
051118 (2007); J. Birjukov, T. Jahnke, and G. Mahler, Eur. Phys.
J. B {\bf 64}, 105 (2008); H. T. Quan, Phys. Rev. E {\bf 79},
041129 (2009); A. M. Zagoskin, S. Savel\'{e}v, F. Nori, and F. V.
Kusmartsev, Phys. Rev. B {\bf 86}, 014501 (2012); O. Abah, J.
Ro\ss nagel, G. Jacob, S. Deffner, F. Schmidt-Kaler, K. Singer,
and E. Lutz, Phys. Rev. Lett. {\bf 109}, 203006 (2012). J. Ro\ss
nagel, O. Abah, F. Schmidt-Kaler, K. Singer, and E. Lutz, Phys.
Rev. Lett. {\bf 112}, 030602 (2014).


\bibitem{kosloff00} R. Kosloff and T. Feldmann, Phys. Rev. E {\bf 61}, 4774
(2000).

\bibitem{kosloffMANY} R. Kosloff and T. Feldmann, Phys. Rev. E {\bf 65}  055102 (R) (2002),
R. Kosloff and T. Feldmann, Phys. Rev. E {\bf 68}  016101 (2003),
T. Feldmann and R. Kosloff, Phys. Rev. E {\bf 70}, 046110 (2004),
T. Feldmann and R. Kosloff, Phys. Rev. E {\bf 85}, 051114 (2012).
\bibitem{wang12} J. Wang, Z. Wu, J. He, Phys. Rev. E {\bf 85}, 041148
(2012).
\bibitem{wang} R. Wang, J. Wang, J. He and Y. Ma, Phys. Rev. E {\bf 87}, 042119 (2013).

\bibitem{nori} H. T. Quan, Yu-Xi Liu, C. P. Sun, F. Nori, Phys. Rev. E {\bf 76}, 031105 (2007).

\bibitem{donor} See also A. E. Allahverdyan, Th. M. Nieuwenhuizen,
Phys. Rev. E {\bf 71}, 046107 (2005).

\bibitem{toy}
S. Toyabe, T. Sagawa, M. Ueda, E. Muneyuki, and M. Sano, Nat.
Phys. {\bf 6}, 988 (2010).

\bibitem{john}
M. Esposito and C. Van den Broeck, Europhys. Lett. {\bf 95}, 40004
(2011); S. Deffner and C. Jarzynski, Phys. Rev. X {\bf 3}, 041003
(2013); J. Goold, U. Poschinger, K. Modi, arXiv:1401.4088 (2014);
J. Goold, M. Paternostro, K. Modi,  arXiv:1402.4499 (2014).

\bibitem{footnote2}
We notice that Eq.(\ref{difirfr}) is equivalent to two explicit
relationships among relative entropies, which read
\begin{eqnarray} && T_B D(\rho_{\tau}|| \rho_B) - T_A
D(\rho_{\tau} || \rho_A ) = T_B D(\rho_{A} || \rho_B )
\label{primad} \\ && \quad = - T_A D(\rho_{B} || \rho_A ) +
(S_A-S_B)(T_A-T_B)\, . \label{secondad}\end{eqnarray} These two
equalities hold for every process $\lambda_i \rightarrow
\lambda_f$ bringing $\rho_i$ into $\rho_{\tau}$; that is, they
hold for every final state $\rho_{\tau}$. They are easily proven
as each of the three sides entering Eqs. (\ref{primad}) and
(\ref{secondad}) is equal to ${\cal U}_A - {\cal U}_B + T_B(S_B -
S_A)$ [which can be demonstrated by using Eq.(\ref{difirfr}) for
the first term and by direct evaluation for the second and third
ones].

\bibitem{tasaki}
H. Tasaki, arXiv::cond-mat/0009244v2 (2000).

\bibitem{footnote3}
A possible derivation of Eq. (\ref{cumu}) goes as follows:

Let $g(x) = \ln (\lavg e^{-x s} \ravg )$ the generating function
for the cumulants, with $C_n = (-1)^n \frac{d^n g}{d x^n} \Bigr
|_{x=0}$. Then, $g(x) = \sum_{n=1} \frac{(-1)^n}{n!} C_n x^n$.
Taking $x=1$, one obtains
$g(1) = \ln \left (\lavg e^{-s} \ravg \right ) = \sum_{n=1} \frac{(-1)^n}{n!} C_n \,
,$ which gives Eq. (\ref{cumu}).

\bibitem{kosloffOSC} Y. Rezek and R. Kosloff, New J. Phys. {\bf 8}, 83 (2006).

\bibitem{kosloffOSC2} E. Torrontegui and R. Kosloff, Phys. Rev. A {\bf 88}, 032103 (2013).

\bibitem{shortcuts}
M. V. Berry, J. Phys. A: Math. Theor. {\bf 42}, 365303 (2009); X.
Chen {\it et al}., Phys. Rev. Lett. {\bf 104}, 063002 (2010);
J.-F. Schaff, X.-L. Song, P. Vignolo, G. Labeyrie, Phys. Rev. A
{\bf 82}, 033430 (2010); J.-F. Schaff, X.-L. Song, P. Capuzzi, P.
Vignolo, G. Labeyrie,  Europhys. Lett. {\bf 93}, 23001 (2011);  M.
G. Bason, M. Viteau, N. Malossi, P. Huillery, E. Arimondo, D.
Ciampini, R. Fazio, V. Giovannetti, R. Mannella and O. Morsch,
Nat. Phys.  {\bf 8} 145 (2012); N. Malossi, M. G. Bason, M.
Viteau, E. Arimondo, R. Mannella, O. Morsch and D. Ciampini, Phys.
Rev. A {\bf 87}, 012116 (2013); A. Del Campo, J. Goold, M.
Paternostro, arXiv:1305.3223 (2013); J. Deng, Q.-h. Wang, Z. Liu,
P. H\"{a}nggi, J. Gong, Phys. Rev. E {\bf 88}, 062122 (2013).

\bibitem{kosloffLUBR} T. Feldmann and R. Kosloff, Phys. Rev. E {\bf 73}, 025107(R) (2006).

\bibitem{messiah}
A. Messiah, Quantum Mechanics, North Holland, Amsterdam (1970).
\end{thebibliography}
\end{document}